# Alternating Current Adiabatic Compression Tokamak:

# A New Way to Fusion Reactor

Yuejiang Shi

*Department of Nuclear Engineering, Seoul National University*

Email: yjshi1973@gmail.com   83565851@qq.com  yjshi@snu.ackr  yjshi@ipp.ac.cn

**Abstract**

*A novel magnetic confined fusion scheme for the fusion reactor is proposed in this paper. A simple numerical estimation shows that ignition of D-T plasma can be achieved in tokamak by major and minor radius adiabatic compression. The density and temperature of compressed plasma can reach $1\text{-}3 \times 10^{22}$ $m^{-3}$ and 15-25keV, respectively. Alternating current adiabatic compression tokamak (ACACT) with $R_0$ = 1.8m can deliver 400MW fusion power with 10% duty cycle of compression pulse.*

**Introduction**

Significant progress has been obtained in magnetic confinement fusion research based on Tokamak type devices in the past 50 years since the first stable and reliable high temperature plasma was achieved in T-3 tokamak [1-2]. The superconducting tokamaks with normal aspect ratio and moderate magnetic field, such as EAST[3], KSTAR[4], JT-60SA[5], and ITER[6], represent the main developing direction at present and in the foreseeable future. Moreover, some compact tokamak types based on high field magnetic field[7-8] or spherical torus[9-10] have also been proposed as the candidate of fusion reactor. One of the main common features of these advanced developed or developing tokamaks is long-pulse steady-state operation.

In this article, an alternating current tokamak with adiabatic magnetic compression pulse (ACACT) is proposed for fusion reactor. The main idea of ACACT is that the extreme high density ignited plasma with Ohmic heating (OH) will be achieved with two step compression. The fusion power after compression can be amplified hundreds times. Considerable fusion energy with moderate fusion energy gain factor can be released in a short compress pulse cycle.

The concept and theory of adiabatic magnetic compression for tokamak plasma was first proposed by Furth and Yoshikawa[11]. Both temperature and density can increase significantly if the compression time is much fast than the energy confinement time. Adiabatic compression experiments have been performed in several tokamaks (ATC[12-14], TUMAN-2[15], TUMAN-3[16], TOSCA[17], TFTR[18], and JET[19]). The experimental results of adiabatic compression in these devices show that the increment of temperature and density are following the adiabatic theory of Furth and Yoshikawa, which verified the possibility and feasibility of adiabatic compression for fusion reactor.

On the other hand, alternating current operation of tokamak has also been demonstrated on quite a few devices (STOR-1M[20], JET[21], CSTN-AC[22], CT-6B[23], ISSTOK[24], and HT-7[25]). Especially, continues waveform (CW) AC operation with 50 discharge cycles and 50s plasma duration was achieved in HT-7.

**Basic equations for adiabatic compression and plasma parameters**

Here, the adiabatic scaling law of Furth and Yoshikawa is briefly cited as following. Neglecting the resistance of hot plasma, on the time scale of the compression, there are two basic constraints:

$$a^2 B_t = \text{const} \quad (1)$$

where $B_t$ is the intensity of toroidal magnetic field (TF), and a is the minor radius of tokamak. For collisional compression ($\gamma=5/3$) we also have the constraint on plasma density and temperature:

$$T n^{-2/3} = \text{const} \quad (2)$$

where $T$ and $n$ are temperature and density, respectively. In terms of the minor and major radii of the tokamak, the compression scaling laws as following:

$$n \propto a^{-2} R^{-1} \quad (3)$$

$$T \propto a^{-4/3} R^{-2/3} \quad (4)$$

$$B_t \propto a^{-2} \quad (5)$$

$$I_p \propto R^{-1} \quad (6)$$

Where $R$ is major radius. $I_p$ is the plasma current. Here, two compression factors are defined as following:

$$C_R = R_0 / R_1 \quad (7)$$

$$C_a = a_0 / a_1 \quad (8)$$

Where the values with subscripts 0 and 1 represent those before compression and after compression, respectively.

According to the above equations, the parameters after compression are as following:

$$n_1 = C_a^2 \, C_R \, n_0 \quad (9)$$

$$T_1 = C_a^{4/3} \, C_R^{2/3} \, T_0 \quad (10)$$

$$B_{t1} = C_a^2 \, B_{t0}, \text{ at plasma magnetic axis} \quad (11)$$

$$B_{t1}=C_a^2 \ C_R^{-1}B_{t0}, \text{ at fixed point} \quad (12)$$

$$I_1=C_R \ I_0 \quad (13)$$

In the ACACT scenario, both minor radius *a* and major radius *R* will be compressed. For simple physics understanding, the compressed process of ACACT is divided two steps. The first step (phase I) is minor radius compression by increasing $B_t$ and keep major radius constant. For this step,

$C_{aI}=a_0/a_I \quad (13)$

$C_{RI}=1 \quad (14)$

$B_{tI}=C_a^2 \ B_{t0} \quad (15)$

$n_I=C_{aI}^2 \ n_0 \quad (16)$

$T_I=C_{aI}^{4/3} \ T_0 \quad (17)$

The second step (phase II) is both major and minor radius compressing by increasing the vertical magnetic field and keep aspect ratio as constant. For this phase,

$C_{RII}=R_{II}/R_0 \quad (18)$

$C_{aII}=C_{RII} \quad (19)$

$n_{II}=C_{RII}^3 \ n_I \quad (20)$

$T_{II}=C_{RII}^2 \ T_I \quad (21)$

So, the final density and temperature after two step compression is as following,

$n_{II}=C_{RII}^3 \ C_{aI}^2 n_0 \quad (22)$

$T_{II}=C_{RII}^2 \ C_{aI}^{4/3} \ T_0 \quad (23)$

The fusion power $P_{DT}$ from neutron produced by D-T reaction with 50%-50% D-T mixture is,

$P_{DT} = 0.56 \times 10^8 \ n^2 V \langle \sigma v \rangle_{DT} \quad (24)$

The unit of $P_{DT}$ and n are GW and $10^{20}m^{-3}$. V is the plasma volume. $\langle \sigma v \rangle_{DT}$ is the reaction rate averaged over Maxwellian distributions, which is related to plasma temperature. Here, the fusion power gain **$P_{DT\text{-gain}}$** is defined as,

$P_{DT\text{-gain}} = P_{DT}^{after \ compression} / P_{DT}^{before \ compress} \quad (25)$

Due to particle number conservation,

$$P_{DT\text{-}gain} \propto n^{\text{afer compression}}/n^{\text{before compress}} = C_{RII}^3 \, C_{aI}^2 \quad (26)$$

Compared to the initial uncompressed plasma, the fusion power of compressed can increase hundreds times with limited compression ratio. For the ACACT operation mode, the initial toroidal field $B_{T0}$ is produced by superconducting TF coils and keep constant during whole discharge. The ideal discharge of ACACT is continues waveform (CW) operation mode which is similar to HT-7. Pulsed adiabatic magnetic compression will be applied in the middle of each discharge cycle. The waveform of two discharge circles is show in fig.1.

The 0-D numerical analysis in following section will be performed to show whether the ignited plasma can be achieved in ACACT. We start from initial phase before compression. The highest density of tokamak plasma is determined by Greenwald density limits[26],

$$\bar{n}_{G20} = \frac{I_p}{\pi a b} = \frac{I_p}{\pi \kappa a^2} \quad (27)$$

$\bar{n}_{G20}$ is in $10^{20}\text{m}^{-3}$, $I_{MA}$ is in MA, $a$ and $b$ is in meter.

And Elongation $\kappa = \frac{b}{a}$ (28)

We take 80% $\bar{n}_{G20}$ for ACACT. The maximum Ip is limited by the minimum edge safety factor $q_{edge}$. Although, many tokamaks demonstrate the low $q_{edge}$ ($q_{edge}<2.5$) discharge, the stable plasma to avoid major disruption in tokamak can be obtained when $q_{edge} >3$. For ACACT, the $q_{edge}$ for initial uncompressed plasma is set as 3. The relation between plasma current and other parameters is as following[27]:

$$I_p = \frac{5a^2 B_{t0}}{q_{edge} R_0} \frac{1+\kappa^2}{2} \quad (29)$$

So eq.27 can be rewritten as

$$\bar{n}_{G20} = \frac{5 B_{t0}}{\pi q_{edge} R_0} \frac{1+\kappa^2}{2\kappa} \quad (30)$$

At the same time, the volume average temperature by OH heating is[28],

$$\langle T_0 \rangle = \frac{(Z_{eff} I_p)^{1/2}}{10 a_0} \quad (31)$$

$<T_0>$ is the volume average density in keV. $Z_{eff}$ is effective charge number.

**Results and discussion**

The averaged density is set as high as 80% Greenwald density. $Z_{eff}$ and $\kappa$ are set as 4 and 1.8, respectively. The initial toroidal magnetic field ($B_{T0}$) and aspect ratio ($\in = R/a$) before compression are assumed as 2T and 3.3, respectively. The relation between the main plasma 5parameters and major radius before compression with OH heating is shown in fig.2. It can be seen that both temperature and density decrease with major radius. We assume that the toroidal field at $R_0$ during compression phase can reach 10T and major radius can be compressed 4 times, which means compression factors are $C_{aI}=5^{1/2}$ and $C_{RII}=4$.

The main parameters after compression are shown in fig.3. It can be seen from fig.3 that the fusion power is not sensitive to machine size ($P_{DT}=3.2GW@R_0=1m$ and $P_{DT}=4.2GW@R_0=1m$), which means that small machine size for ACACT is reasonable. On the other hand, the smallest of ACACT is limited by the Lawson criterion which will determine whether the reactor can be ignited. The triple production for ignition condition for D-T reaction is[29],

$nT\tau_E > 3\times10^{21}$ $m^{-3}keVs$  (32)

$\tau_E$ is the energy confinement time. Compared to other fusion devices, one advantage of tokamak is that there are abundant experimental data and relative mature transport theory to support some reliable scaling law. The energy confinement time can be estimated for the proposed tokamaks or some new machine under construction with these scaling laws. Fig.4 shows the energy confinement time for the compressed plasma of ACACT with some well-known scaling laws for the energy confinement time $\tau_E$ of tokamak plasmas (Goldston[30], Kaye-All and Kaye-Big[31], ITER89P and ITER89OL[32], ITER93HP[33], ITER98P, ITER98HP$_{ELMy}$ and ITER98HP$_{ELMy\ free}$ [34]). All these scaling laws predict the energy confinement time increase with machine size. At the same time, the scaling law of ITER89P for L-mode plasma gives the worst prediction of $\tau_E$. On the other hand, $\tau_E$ of ITER98HP$_{ELMy\ free}$ for ELMy free H-mode plasma has the highest expected value. So the triple productions from ITER89P and ITER98HP$_{ELMy\ free}$ in fig.3 represent the worst predicted value and best predicted value for ACACT. According to scaling law of ITER89P, the initial major radius of ACACT should be larger than 1.5 m to achieve ignited plasma. On the other hand, the Lawson criterion from the prediction of ITER98HP can be achieved for smaller machine size of ACACT. Anyway, the ignited plasma can be obtained for ACACT with big enough size.

While, the Lawson criterion is just one necessary requirement, not the complete restrictions for the fusion reactor. The power or energy gain, $Q_{DT}$, is the basic factor to determine the economic feasibility of one fusion reactor. Generally, $Q_{DT}$ for tokamak is defined as the ration of neutron power to the heating power. However, such $Q_{DT}$ in physics concept is meaningful for long-pulse steady state tokamak, not practical for the pulsed operated machine such as ACACT and inertial confined devices. Here, the effective $Q_{DT}$ is defined as following,

$Q_{DT}^{effective} = \dfrac{E_{DT}}{E_{in}}$ (33)

where $E_{DT}$ is the total energy carried by the neutrons produced by D-T reaction during one compression pulse. Compared to $Q_{DT}$, $Q_{DT}^{effective}$ is more close to the actual 'Engineering' value.

$E_{in}$ is the total electrical energy to make one discharge cycles of ACACT. It is well known that most of $E_{in}$ for tokamak is used to charge the magnetic coils, especially the PF coils. For one medium size tokamak with R≈1.8m, the consumed electrical energy to charge magnetic coils to produce and sustain 1MA plasma current for several seconds is about 50MJ ~100MJ. If the electrical energy used for compression process is 100MJ, the total consumed electrical energy for one ACACT discharge cycle will be 150MJ-250MJ. The fusion neutron power $P_{DT}$ for ACACT@ $R_0$=1.8m (as shown in fig.3) is about 4GW. Then $Q_{DT}^{effective}$ can reach 5 with the 0.19s~0.31s compression pulse. On the other hand, the effective output power of ACACT is determined by the ratio of compression pulse duration to the total discharge duration as following,

$$P_{DT}^{effective} = P_{DT} \times \delta_{compress} \quad (34)$$

where $\delta_{compress}$ is the duty cycle of compression pulse. The effective power of ACACT with $R_0$=1.8m is about 400MW (4GW×10%) if the duty cycle of compression pulse is 10%. On the other hand, the fusion power of ITER with R=6.2m will be 500MW-700MW. Although the volume size of ITER is around 40 times as ACACT, the fusion power of these two machines is quite close to each other due to the extremely high density plasma of ACACT.

**Methods**

The concept sketch of cross section of ACACT is shown in Fig.5. Although ACACT is not a steady-state operated machine, the primary toroidal magnetic field to confine the uncompressed plasma needs to sustain long pulse to fulfil the request of continues waveform (CW) AC operation. Superconductor is the best choice for the primary 2T TF coil. However, the conditional low temperature superconductor cooled by liquid helium is not suitable for ACACT due to the limited space in the center of ACACT. The developing high temperature (HTC) superconductor will be a good solution. On the other hand, conventional conductor will be applied for the secondary 8T TF coil for compression pulse because the ramping-up rate of magnetic field of superconductor is too slow to achieve the requirement of adiabatic compression. The PF coils involved in shaping and compression processes should also be made by conventional conductor.

In summary, the physics feasibility for ACACT is analyzed in this article. 0D numerical simulation shows that ignition of D-T plasma can be achieved in a small ACACT tokamak. The main bottleneck to prevent the target of ACACT is some engineering technology, especially the challenging requirement for magnetic conductor of ACACT. Anyway, the breakthrough in engineering technology is very important and will definitely accelerate the realization of fusion power plant.

**Acknowledgement**

This research was supported by Basic Science Research Program through the National Research Foundation of Korea (NRF) funded by the Ministry of Science and ICT (No. NRF-2018R1A2B2008692).

**Figures**

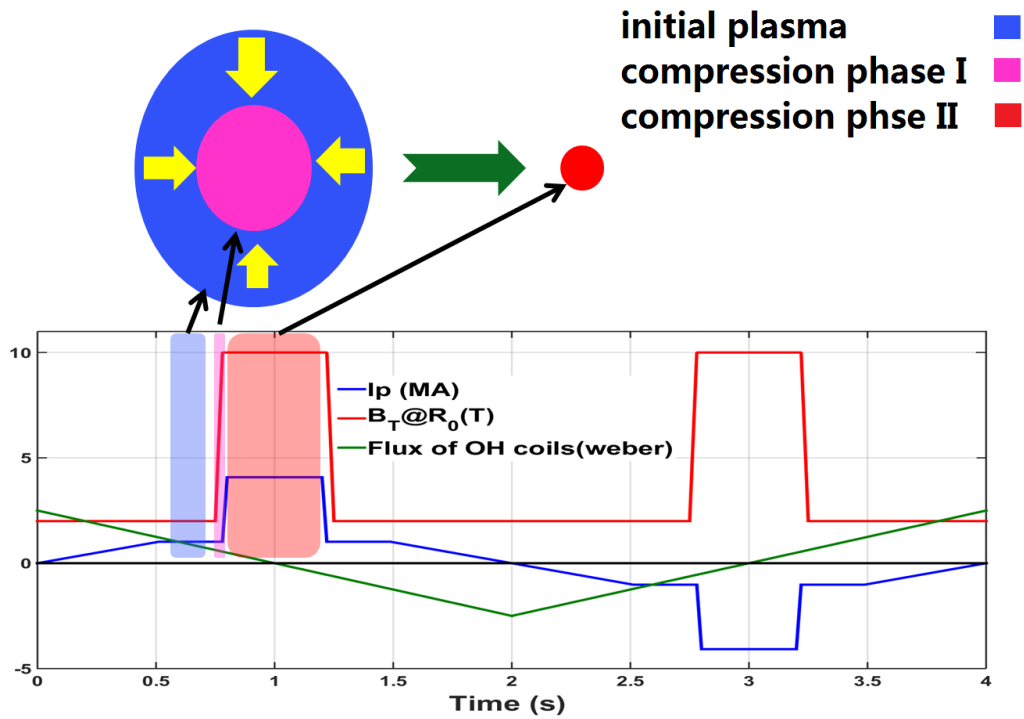

Fig.1  Waveform of two discharge cycles of ACACT.

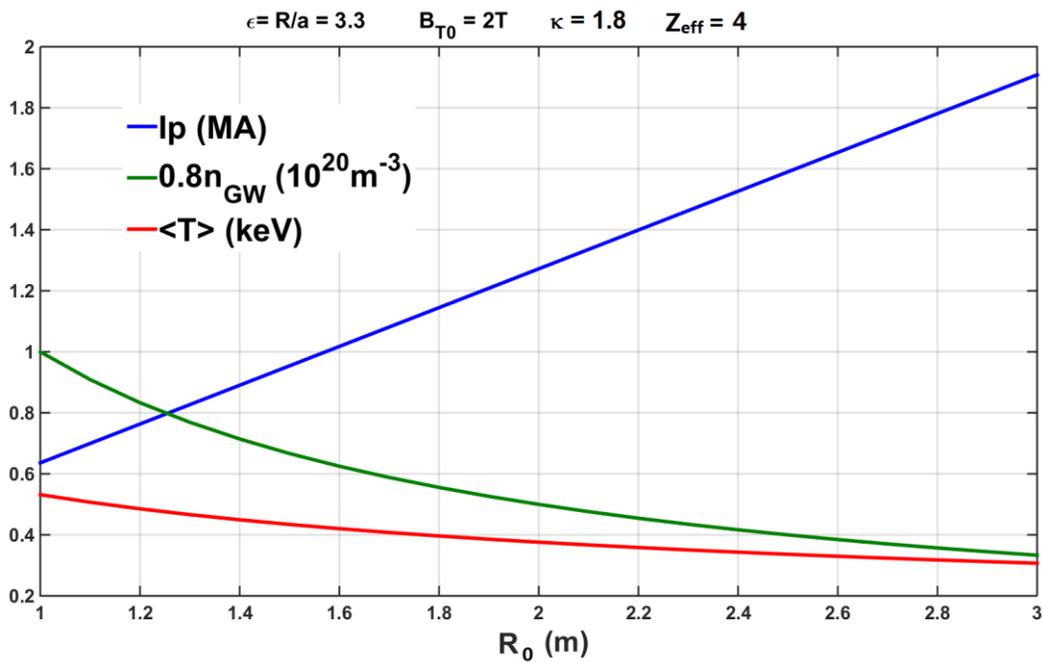

Fig.2  The main parameters of the initial uncompressed plasma with OH heating in ACACT.

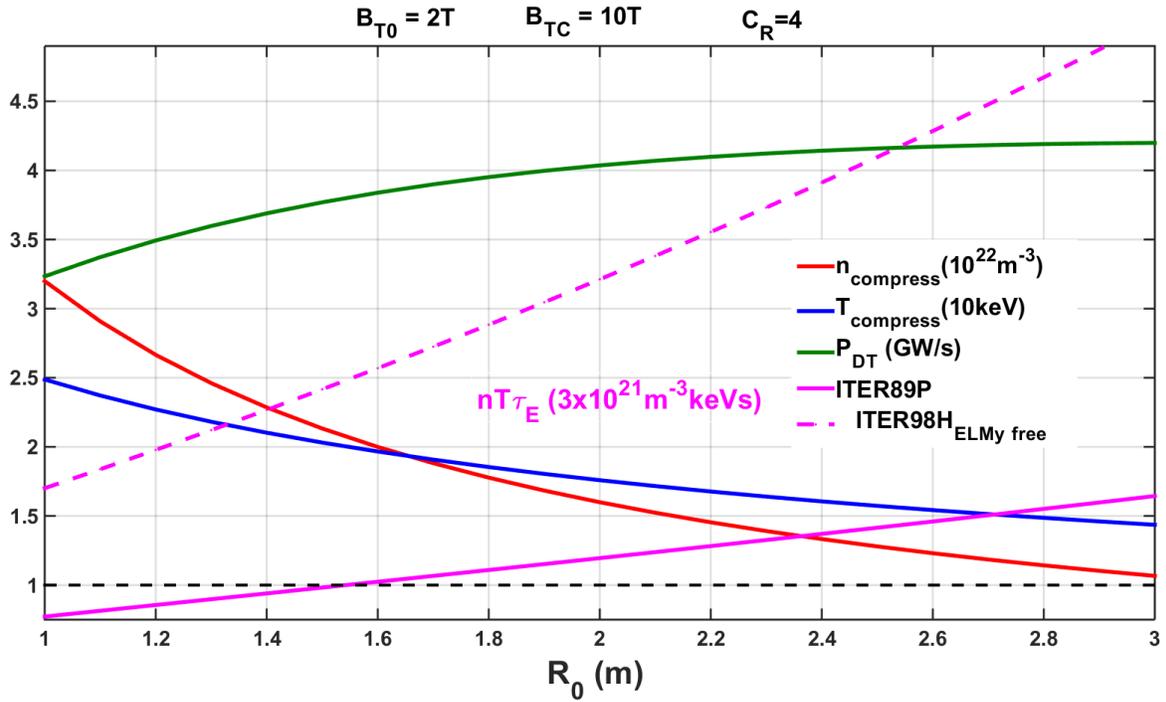

Fig.3  The main parameters of the compressed plasma in ACACT.

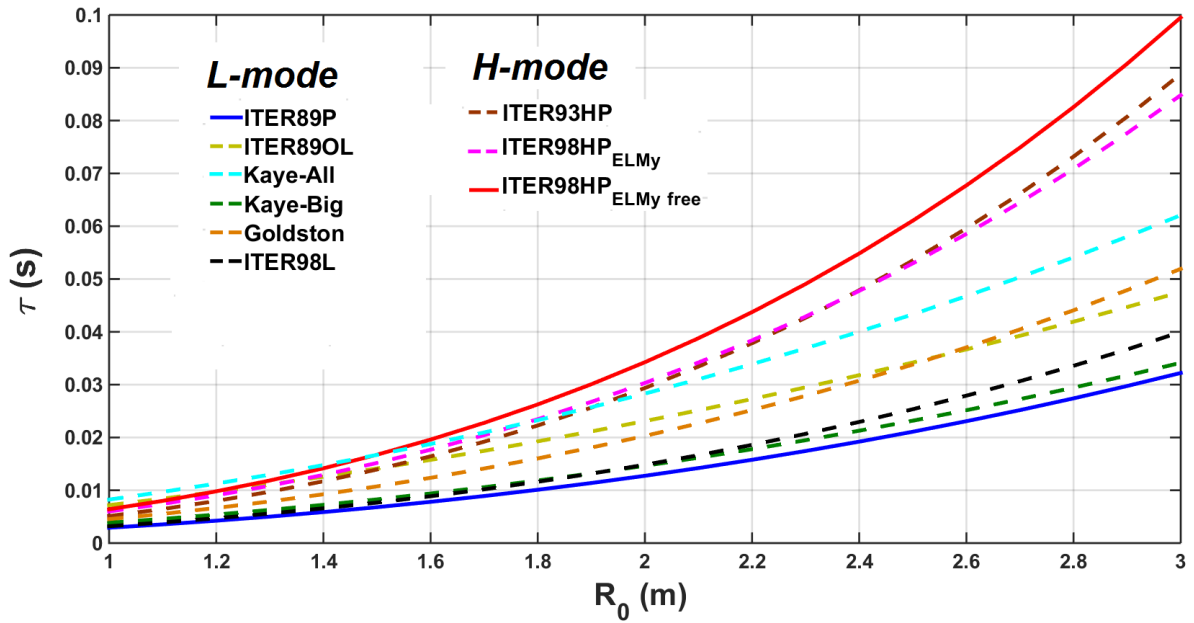

Fig.4  The energy confinement time for the compressed plasma in ACACT.

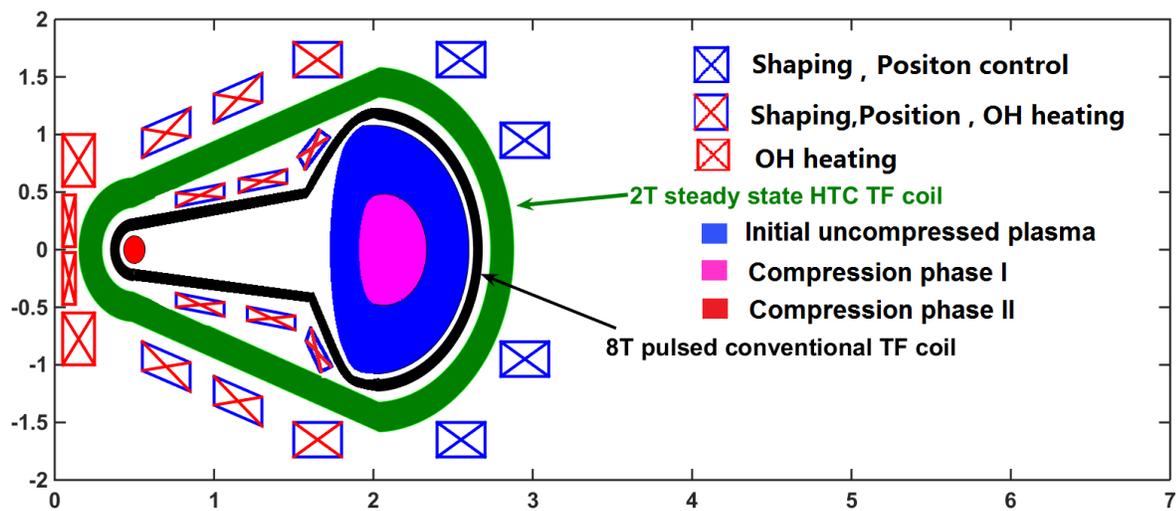

Fig.5  The concept sketch of cross section of ACACT.